\documentclass[preprint2]{aastex}

\shortauthors{Monta\~n\'es-Rodr\'{\i}guez et al., 2005}

\begin{document}

\title{Vegetation signature in the observed globally-integrated
spectrum of Earth: Modeling the red-edge strength using simultaneous
cloud data and application for extrasolar planets}

\author{P. Monta\~n\'es-Rodr\'{\i}guez\altaffilmark{1}, E. Pall\'e\altaffilmark{1} and P.R. Goode\altaffilmark{1}}
\affil{Big Bear Solar Observatory, New Jersey Institute of Technology,
Newark, NJ 07102, USA}
\email{pmr@bbso.njit.edu}




\begin{abstract}

A series of missions will be launched over the next few
decades that will be designed to detect and characterize
extrasolar planets around nearby stars. These missions
will search for habitable environments and signs of life
(biosignatures) in planetary spectra. The vegetation's 
``red edge'', an intensity bump in the Earth's spectrum
near 700 $nm$ when sunlight is reflected from greenery, is
often suggested as a tool in the search for life in
terrestrial-like extrasolar planets. Here, through
ground-based observations of the Earth's spectrum,
satellite observations of clouds, and an advanced
atmospheric radiative transfer code, we determine the
temporal evolution of the vegetation signature of Earth.
We find a strong correlation between the evolution of the
spectral intensity of the red edge and changes in the
cloud-free vegetated area over the course of the
observations. This relative increase for our single day
corresponds to an apparent reflectance change of about
0.0050$\pm$0.0005, with respect to the mean albedo of 0.25
at 680 $nm$ (2.0$\pm$0.2\%). The excellent agreement
between models and observations motivated us to probe more
deeply into the red edge detectability using real cloud
observations at longer time scales. Overall, we find the
evolution of the red edge signal in the globally-averaged
spectra to be weak, and only attributable to vegetation
changes when the real land and cloud distributions for the
day are known. However, it becomes prominent under certain
Sun-Earth-Moon orbital geometries, which are applicable to
the search for life in extrasolar planets. Our results
indicate that vegetation detection in Earth-like planets
will require a considerable level of instrumental
precision and will be a difficult task, but not as
difficult as the normally weak earthshine signal might
seem to suggest.
\end{abstract}

\keywords{Earth's albedo, Earthshine spectrum, astrobiology:  biomarkers, red
edge}

\section{Introduction}

A spectacular astronomical revolution is on its way with the
development of new scientific projects aimed at the detection and
characterization of Earth-size extrasolar planets (Fridlung 2004).
Among the most prominent goals of these missions will be the search
for biosignatures indicating the presence of life, but life can
exist in a rich variety of forms, if the Earth's past and present is
to serve us as a guide.

For most of its past history, life on Earth existed solely as
unicellular microorganisms, which were able to interact with, and
transform their environments. In the early Archean period (4.0-2.6
$Gyr$ b.p.) for instance, methanogens were already producing
$CH_{4}$, which also has a small abiotic source, and could have
generated a detectable methane-rich atmosphere (Schindler \& Kasting
2000). With the advent of cyanobacteria and the subsequent rise in
$O_{2}$ concentration, even larger changes in atmospheric
composition took place. Thus, the remote detection of such life
forms might be possible by studying the atmospheric composition of a
planet (Hitchcock \& Lovelock 1967).

Suppose that the necessary physical conditions for life were found on
a planet, and biological fingerprints are detected in its atmosphere
(Hitchcock \& Lovelock 1967; Selsis et al. 2002; DesMarais et al.
2002; Ford et al. 2001). Could we determine whether this life has
evolved further than unicellular organisms into more complex
organisms such as plants? The Earth's vegetation has several
spectral features, related to chlorophyll, that make detection of
plants on Earth an easy task from space when geographical resolution
is available. In particular the `red edge', a sharp increase in leaf
reflectance around 700\,$nm$ (Clark et al., 1993, Kiang et al., 2005), 
was already detected in 1993 by the Galileo mission (Sagan et al. 1993).
Nowadays, vegetation leaf indices and phytoplankton blooms are
routinely monitored from space.

However, when observing an extrasolar planet, all the reflected
starlight and radiated emission from its surface and atmosphere will
be integrated into a single spectrum. A similar kind of
globally-integrated planetary spectrum for the Earth can be measured
by observing the earthshine (Qiu et al. 2003; Pall\'e et al. 2003;
Pall\'e et al. 2004), the light reflected by the daytime Earth onto
the dark portion of the lunar surface.

Spectral measurements of the earthshine are one of the few
observations one can make of disk-integrated Earth spectra. These
earthshine measurements allow us to observe the Earth as it would be
seen from a distant planet, i.e., with no spatial resolution. With
the recent successes in the detection of exoplanets, the
characterization of the spatially averaged spectral reflectance of
the Earth, as provided by the earthshine technique, is becoming a
keystone in the search for habitable exoplanets.

Several groups have reported in the literature analysis of their
observations (Woolf et al. 2002; Arnold et al. 2002;
Monta\~n\'es-Rodr\'iguez et al. 2004; Monta\~n\'es-Rodr\'iguez et
al. 2005; Seager et al. 2005, Hamdani et al., 2006) or modeling
(Arnold et al. 2002; Tinetti et al. 2006a) of the earthshine and
tried to determine the detectability of the red edge. While some
authors claimed to have detected the red edge, others are doubtful
and the overall agreement between earthshine modeling and
observations has been, so far, limited. The ever-present blocking
cloudiness on global scales, with its high albedo, would be expected
to obscure the red edge signal.

In a previous work (Monta\~n\'es-Rodr\'iguez et al., 2005), we
measured with great precision the spectral albedo of the sunlit
Earth, $p^*(\lambda)$, as reflected from the Moon on 2003 November
19. The earthshine-contributing area during the observations was
centered sequentially over Western Africa, the Atlantic Ocean and
the Amazonian rainforest. Our analysis allowed us to determine the
scale of $p^*(\lambda)$, which was comparable to observations of the
photometric albedo for the subsequent day, and had an average value
of 0.27$\pm$0.01. In Monta\~n\'es-Rodr\'iguez et al. (2005), the
good agreement between the independent photometric and spectroscopic
observations of albedo confirmed the validity of applying our
photometric data reduction methodology (Qiu et al., 2003; Pall\'e et
al., 2003) to spectral data reduction.

Our analysis of earthshine data for 2003 November 19 did not show a
significant vegetation signal in the Earth's globally-integrated
spectrum. We speculated that the lack of a strong red edge was due
to cloud obstruction, because the Earth's albedo is dominated by the
total cloud amount and optical thickness (Pall\'e et al., 2004).
However, we could not compare with the real cloud distribution at
the time of observations because they were not yet available. Since
then, global satellite observations of clouds from the International
Satellite Cloud Climatology Project (ISCCP) have been released 
covering 2003 November 19 in the latest update, which now covers the
period from July 1983 to December 2004. With
these data, we have reproduced the Earth's scene for 2003 November
19, such as it would have been viewed from a lunar perspective, and
we have generated a synthetic spectral reflectance for that date and
time. Finally, we have compared and contrasted our simulations and
observations, and quantified the changing slope of the red edge, as
Earth rotates and cloud-free vegetated areas appear and disappear
from the earthshine.

\begin{figure}[h]
\epsscale{0.85} \plotone{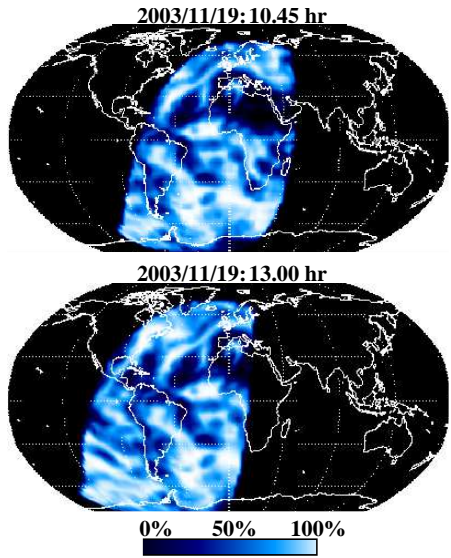} \caption{Map of the
earthshine-contributing area during the observations on 2003
November 19. Top: 10:45 UT; Bottom: 13:00 UT. Although the
earthshine contributing area and solar and lunar cosines have been
calculated for the exact times of our observations, the time
resolution of the ISCCP data is 3 hours. Thus, the cloud data
represented in the two maps correspond to 9:00-12:00 UT and
12:00-15:00 UT ISCCP data bins. Still, one can note how little the
cloud cover did change during the night. In particular, compare the
clouds over South America in the upper and lower panels.}
\label{fig1}
\end{figure}

\section{Data}

\subsection{Earthshine Observations}

Earthshine spectroscopic observations were taken from Palomar
Observatory with the 60-inch telescope (P60").

A single order, long slit spectrograph with an entrance slit of
1.32" x 6' on the sky was selected; for comparison, the mean lunar
apparent diameter during the time of observations was 32.24'. We
covered the spectral region between 460\,nm and 950\,nm with a
resolution of 7.1 \AA at 700\,nm. We took alternate exposures of the
bright and dark side of the Moon, the dark side exposures including
a large portion of the background sky, with signal to noise ratios
of 500 and 30 respectively. The ratio between the signal of the dark
side and the immediately surrounding background sky was of the order
of 2.5. The ratio between dark and bright side spectra gives us,
after accounting for several geometric corrections, the Earth's
albedo as a function of wavelength. Detailed data acquisition
techniques and data reduction for the earthshine spectra presented
here can be found in Monta\~n\'es-Rodr\'iguez et al. (2005).

\subsection{Satellite Cloud data}

 The ISCCP dataset provides continuous global
coverage of cloud amount and radiative properties, from the combined
observations of several inter-calibrated geosynchronous and polar
satellites. In the ISCCP dataset, total cloudiness is determined
using both visible and infrared radiances, whereas the separation
into low, mid and high level cloud types is determined using
infrared radiances only. The data are given for 280x280 km$^2$ cells
with the cloud fraction in each cell determined by dividing the
number of cloudy pixels by the total number of pixels per cell. All
ISCCP data products are available at the ISCCP Central
Archive\footnote{http://isccp.giss.nasa.gov }, and a detailed
description of the whole dataset can be found in Rossow et~al.
(1996).

In Figure~\ref{fig1}, we have used the ISCCP data to determine the
evolution of the cloud-geography pattern for the date (2003 November
19) and times (from 10:28:12.0 to 13:04:47.9 UT) of our
observations. Note that on the timescale of our observations ($\sim$
3 hours), the variability in the large-scale cloud patterns is very
small, and the clouds can be considered fixed with respect to the
geography. We have calculated the percentages of cloud-covered areas
and cloud-free oceans, land, ice and vegetated areas over the entire
earthshine-contributing region during observations (given in Table
1). The cloud percentage at each grid point of the Earth has been
further subdivided into low, mid and high level clouds according to
the ISCCP IR radiance classifications. To simulate the
Sun-Earth-Moon earthshine geometry for the date/times and thus allow
a direct comparison of the model and observations, these global
percentages have been weighted by the solar and lunar cosines at
each point of the Earth.

\begin{table}[h]
\caption{Composition (in percentages) of the Earth scene for 2003
November 19 in the earthshine field of view for our observations.
Clouds refer to the total cloud amount including low, middle and
high altitude clouds. Land percentages includes shrub, tundra and
desert regions free of clouds. Vegetation refers to cloud-free
vegetated areas. The total ice/snow area during observations was
always less than 0.1\%.} \vspace{2mm}
\begin{tabular}{cccccc}
\hline
Time (UT)  & clouds & oceans & land  & vegetation \\
\hline
10:00   & 58.59  &  16.37& 9.03  & 15.90   \\
10:30   & 59.34  &  17.33& 7.90 & 15.34   \\
11:00 & 59.90  &  18.52& 6.76 & 14.75   \\
11:30 & 60.27  &  19.73& 5.70  & 14.23   \\
12:00 & 60.46  &  20.72& 4.75  & 14.01   \\
12:30 & 60.42  &  21.36& 3.99  & 14.16   \\
13:00 & 60.30  &  21.73& 3.36 & 14.56   \\
\hline
\end{tabular}
\label{lunar_ref}
\end{table}

\section{Radiative transfer simulations of the earthshine}

The synthetic generation of earthshine spectra has been carried out
using the state-of-the-art atmospheric code FUTBOLIN. FUTBOLIN (FUll
Transfer By Optimized LINe-by-line methods) is a multi-level,
multiple scattering radiative transfer model for the calculation of
line-by-line atmospheric emission/transmission spectra in planetary
atmospheres (Mart\'in-Torres et al. 2005; Kratz et al. 2005). It can
generate high-resolution synthetic spectra in the 0.3-1000 micron
spectral range, using as its principle inputs the atmospheric
profiles from a 3-D global circulation model (Garc\'ia et al. 1992),
the spectral albedos from the ASTER (Advanced Spaceborne Thermal
Emission and Reflection Radiometer) and TES (Thermal Emission
Spectrometer) libraries\footnote{http://speclib.jpl.nasa.gov ;
http://tes.la.asu.edu }, and the spectroscopic data from the
HITRAN2K (HIgh-resolution TRANsmission molecular absorption)
database (Rothman et al. 2003). Solar irradiance data in the 490-940\,
$nm$ spectral region is taken from the AIRS (Atmospheric InfraRed
Sounder)database.

FUTBOLIN has been previously compared to other atmospheric models,
such as LBRTM, LINEPAK, GENLN2 and MRTA, with a remarkably good
agreement in the results, with differences being of the order of
0.5\% (Kratz et al. 2005). It has been also successfully applied to
the analysis of terrestrial and planetary spectra of Venus and Titan
(Mart\'in-Torres et al. 2003a, and 2003b).

For the simulations shown in this paper, the Earth is treated in 1-D,
as a single cell, for which the averaged atmospheric, surface and
cloud properties are given. The atmosphere for the night was
described through a standard atmospheric composition, temperature
and pressure profiles (Garc\'ia et al. 1992). We have used Voigt
profiles as the lineshapes of atmospheric gases through the
atmosphere, and a number of physical phenomena like multiple
scattering, continuum absorption by water vapor, $CO_{2}$, $N_{2}$
and $O_{2}$, and line mixing in the $CO_{2}$ bands. The spectra have
been computed with a $0.0005 cm^{-1}$ resolution and have been
degraded to the observation's instrumental resolution.

Using global cloud maps from ISCCP, we calculated a global mean
percentage of clouds and surface types, weighted by the solar and
lunar cosines. These percentages were used to average the computed
spectra for each type of surface considered into a global mean.
Clouds were separated into the three ISCCP IR cloud categories (low,
mid and high) have been modeled as Strato-cumulus, Alto-stratus and
Cirrus, respectively (Manabe \& Strickler 1964). Cloud spectral
calculation include multiple scattering (Mie scattering) and are
parameterized using bi-directional reflectance functions (BDRFs) for
each cloud type. The mean solar and lunar cosine angles over the
earthshine-contributing area, were defined as the incident and
reflecting angles respectively.

\begin{figure}
\epsscale{0.95} \plotone{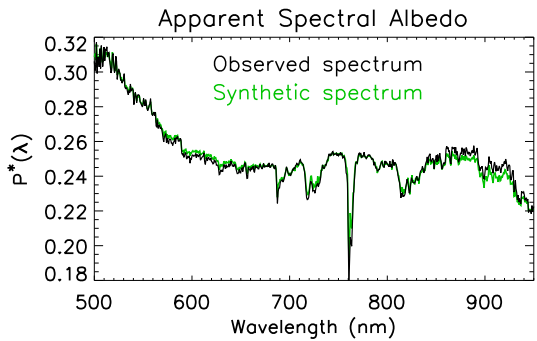} \caption{Averaged observed
and synthetic Earth's apparent spectral albedo for 2003 November 19,
for the entire spectral range of our observations.} \label{fig2}
\end{figure}

The resultant synthetic spectrum for 2003 November 19, based on the
mean of the percentages given in Table 1, is compared to the average
of all observed spectra in Figure~\ref{fig2}. As can be seen in the
figure, there is an excellent agreement over almost the entire
spectral range covered; the major discrepancy is in the depth of the
absorption bands and derives from the complexity of the treatment of
the multiple scattering in the simulations. This agreement
represents a real improvement over earlier efforts to fit earthshine
observations with models, and gives us strong confidence in our
observational data.

\begin{figure*}
\epsscale{1.5} \plotone{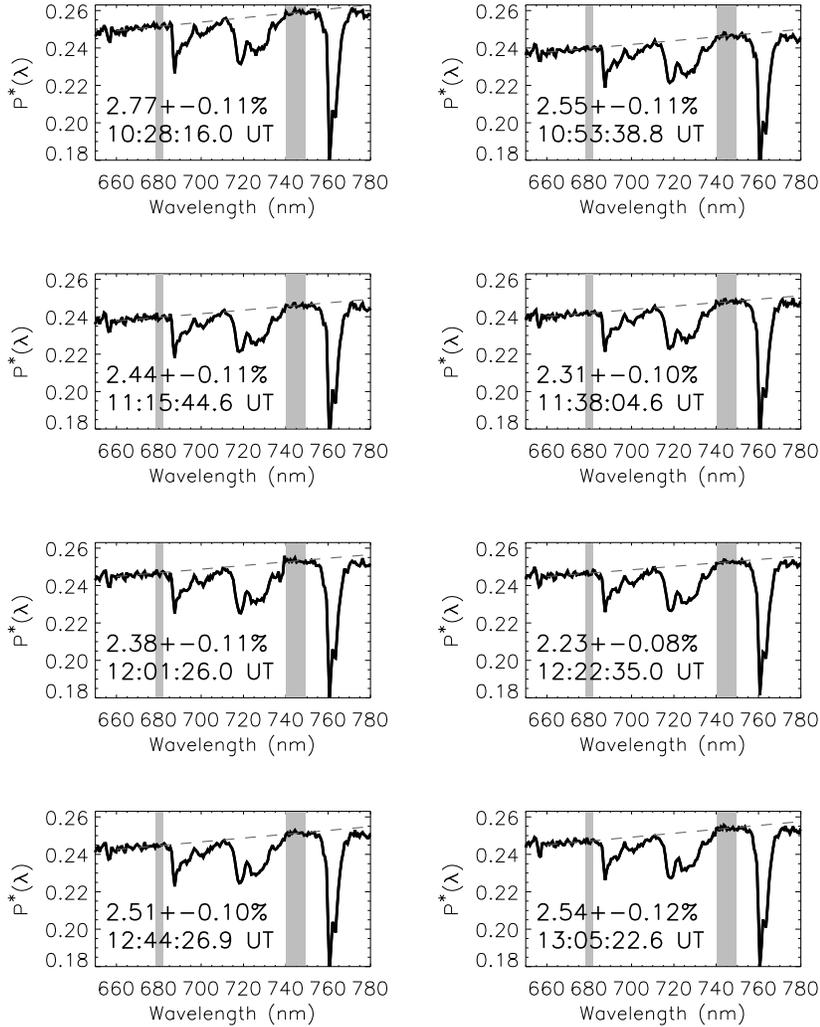} \caption{Evolution of
the earthshine spectrum during the observations on 2003 November 19.
Shadowed strips in each panel indicate the two sectors used to
determine the red-edge slope. The percentage rise between 680 and
740\,nm, and the time of observation (UT) are also indicated. The
narrow strip in each panel is a continuum region that doesn't show
much variation. Various spectral positions and widths of the narrow
band of continuum were tried with similar results for the relative
increase of the red edge bump. The dashed line in each panel
connecting the broad and narrow strips indicates the slope from
which the change in intensity in the red edge is determined.}
\label{fig3}
\end{figure*}

\section{Red edge evolution during observations}

In Figure~\ref{fig3}, the eight earthshine spectra taken on 2003
November 19 are plotted. To determine the intensity of the red edge,
we calculated the slope of a straight line connecting the red edge
to the nearby spectral continuum. This is the dashed line, in each
of the eight panels of Figure~\ref{fig3}, connecting two selected
spectral regions (marked as shadowed strips also in
Figure~\ref{fig3}). The choice of bands is made to avoid the
atmospheric absorption bands of O$_2$-B (centered at 690\,nm,),
H$_2$O (centered at 720\,nm) and O$_2$-A (centered at 760\,nm). The
regions were fixed to the ranges 678--682\,nm and 740--750\,nm and
contained 6 and 14 data points, respectively. We determined the
percentage increases of the spectra between 680 and 740\,nm, by
applying a linear fit to all data points contained within the two
selected regions, for each of the eight observed spectra and for a
set of seven synthetic spectra. The slope of the dashed line was
relatively insensitive to the choice of the narrow band near 680\,
$nm$ in Figure~\ref{fig3}. That is, the relative slopes of the red
edge where about the same whether the band near 680\,$nm$ was made
broader or moved to a slightly different wavelength. The synthetic
spectra were generated with a time step of 30 minutes, using the
percentages given in Table 1, and also distinguishing between low,
middle and high clouds and a variety of land surfaces (shrubs,
tundra and deserts).

The percent increase in reflectance for the observed spectra varied
between 2.77$\pm$0.22\% and 2.23$\pm$0.16\%, and are indicated in
Figure~\ref{fig3}. In the case of the synthetic spectra, the slopes
were consistent with, but more subtle than the variations in the
observational data.  In detail, the percent increases in the models
varied between 0.30$\pm$0.01\% and 0.07$\pm$0.01\%. We attribute the
more subtle model variations to the difficulty in the treatment of the
scattering. In particular, the continuum near 680\,$nm$
is inside a weak $O_{3}$ band, and we anticipate that this part of
the spectrum would drop with an improved treatment of the multiple
scattering, thus improving the agreement with the observed slopes in
Figure~\ref{fig3}. The rise in the observed spectra was in general
more sensitive to the width of the spectral regions selected for the
linear fit, than it was in the synthetic spectra due to the noise in
the observational data.


\begin{figure}[h]
\epsscale{1.05} \plotone{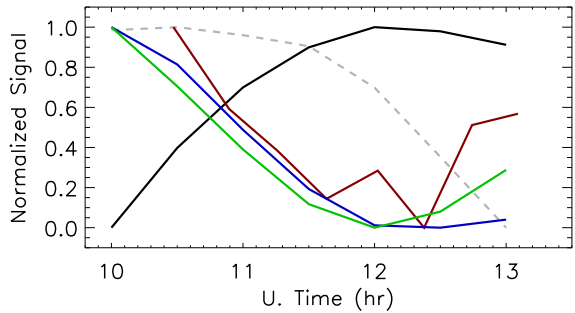} \caption{Evolution of
different signals during observations: 680--740\,nm rise for the
models (blue). 680--740\,nm rise for the observations (red). Total
cloud-free vegetated areas (green). Total-cloud amount (solid gray).
Low-cloud amount (dotted gray).} \label{fig4}
\end{figure}

\begin{figure}[h]
\epsscale{1.05} \plotone{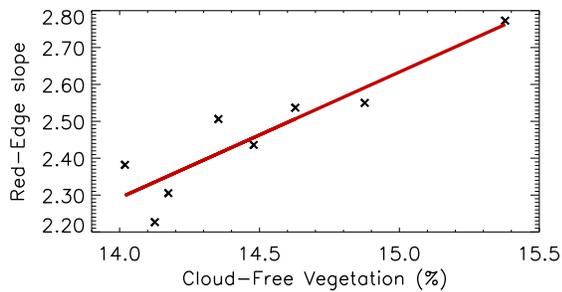} \caption{Scatter plot of
the projected cloud-free vegetated areas measured from satellite,
and the strength of the red edge slope in the earthshine
observations.} \label{fig4b}
\end{figure}

We emphasize that for both models and observations, the slopes
varied according to the change in the percentage of cloud-free
vegetated areas, and in opposition (anti-correlation) to the change
in total cloud amount. The normalized evolution of the cloud-free
vegetated areas, cloud coverage and the red-edge rises are shown in
Figure~\ref{fig4}. We calculated the correlation between the decline
of normalized cloud-free areas of vegetation and the red-edge
decline for models and observations over the course of the
observations. To correlate the areas of vegetation with the rise of
the red edge in the observations, the former were interpolated to
the times of our observations. The correlations were 0.95 for the
models and 0.91 for the observations, respectively. In
Figure~\ref{fig4b} the observed red edge slope in our earthshine
measurements is plotted against the cloud-free vegetated area
measured from satellite data. The significant correlation shown in
the plot is probably the first unambiguous detection of vegetation
on Earth on global scales.

However, we note that vegetation is not the only possible source of the bump
in the 680--740\,nm region, clouds effects may also cause it because
varying the quantity or type of clouds into the scenery will produce
differential changes in albedo along the spectra (Tinetti et al.,
2006a). In particular, low clouds are bright at these wavelengths
and add to the red-edge slope, while other relatively darker surface
components, such as oceans, non-vegetated land areas and snow or
ice, cause a decline in the red edge intensity (Tinetti et al.,
2006a). Thus, the quantification of the red-edge slope does not
alone provide a direct vegetation strength index in
globally-integrated measurements. However, we find no significant
correlation between the low cloud amount and the red edge
variability during our observations (see Figure~\ref{fig4}).

To eliminate the possibility that the decrease found in the red-edge
slope was due to changes in cloud percentage in the evolving field
of view during the observations, we carried out a final test. New
models were calculated for the times shown in Table 1. All of them
with a fixed cloud amount in the field of view, the same found at
the beginning of our observations. The same percentages shown in
Table 1 were used for the remaining components of the scene, except
for the oceans. The percentage of total ocean area was slightly
modified to ensure that the total of all components add to 100\,\%
and compensate for the fixed cloud percentages. The correlation
found between the red-edge rises for these new models and the
evolution of the cloud-free vegetated areas was still 0.93,
indicating that the cloud-free vegetated areas, and not clouds, are
responsible of the red-edge evolution seen in our observations.

Thus, we conclude that the observed evolution of the red edge signal
during our observations is attributable to the change in cloud-free
vegetated areas contributing to the earthshine at each time.
However, this attribution is only possible because the real land and
cloud distributions for the day are known. Without these data, which
will not be available in the case of an extrasolar planet
observation, we could not conclude that we are detecting the red edge 
(Monta\~n\'es-Rodr\'iguez et al, 2005).

\section{Evolution of the red edge at longer time scales}

\begin{figure}
\epsscale{0.9} \plotone{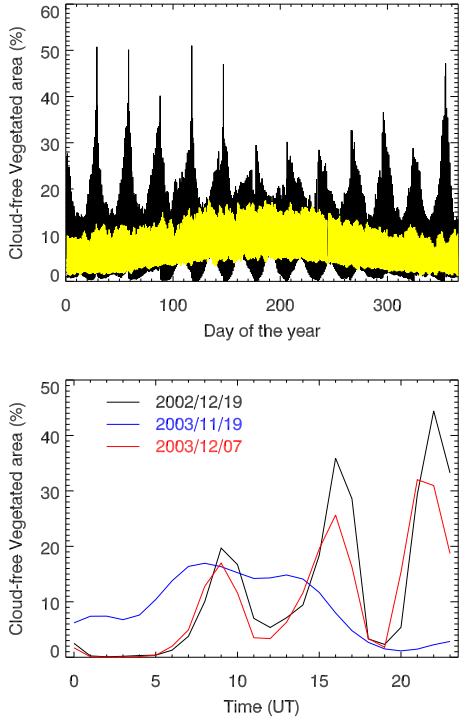} \caption{Top
panel: Percentage of the total earthshine-contributing area that is
simultaneously cloud-free and covered by vegetation (black) though
the year 2003. The percentages are weighted by the solar and lunar
cosines and calculated with a 1-hour time resolution. In yellow are
the percentages of total sunlit area which are simultaneously
cloud-free and covered by vegetation. Note the largest excursion of
the black curve due to the Moon's orbital cycle (glint scattering of
sunlight from a heavily vegetated areas of a crescent Earth) .
Bottom panel: Top panel data for three particular days, but showing
the part of the earthshine heading toward the Moon. Blue: 2003
November 19 a day with an average vegetation percentage
contribution. Black: 2002 December 19 a day with extreme vegetation
percentage (and with an earthshine-contributing area near zero).
Red: 2003 December 7, a day with a large, detectable vegetation
percentage (30\%) and an earthshine-contributing area 10\% of the
maximum (1/2 the total of the Earth's surface).  The three peaks in
the red and black curves correspond to times when heavily vegetated
areas rotate into view (in this case: Asia, Africa and South America
from left to right).} \label{fig5}
\end{figure}

The excellent agreement between our observations and the models,
motivated us to study the evolution of the vegetation signature in
the Earth's spectra on larger time scales, for which observations
are not available but global cloud data is.

Here we have modeled the temporal evolution of the vegetation's
signal strength in the earthshine using cloud cover maps from the
ISCCP dataset. Considering the relative positions of the
Sun-Earth-Moon system, the percentage of the Earth's surface that is
free of clouds and visible from the Moon is computed for each day at
each point of a grid surface. This percentage is subdivided into
percentages corresponding to oceans, snow/ice, vegetated areas,
shrubs, deserts and tundra. Similarly the percentages of low, mid
and high-altitude clouds are computed. These percentages are then
combined for the whole earthshine-contributing area, weighted by the
solar and lunar cosines at each grid point to account for the true
earthshine signal. For comparison, we have also calculated these
percentages in the direction of the Sun (retroflection) from where
the whole sunlit half of the Earth's surface is always visible. Our
projected percentages are used as input for FUTBOLIN, which
calculates the spectral features of such a projection.

\begin{figure}[h]
\epsscale{0.9} \plotone{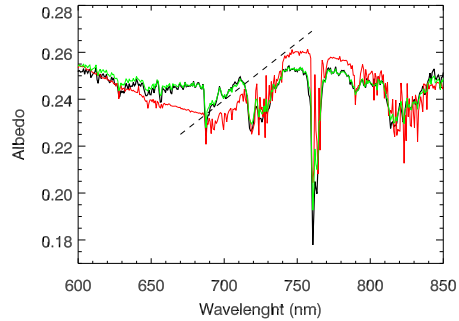} \caption{Black:
Earthshine spectral albedo measured from Palomar Mountain
observatory on 19 November 2003. Blue: Modeled spectral albedo using
the FUTBOLIN code and the weighted percentage coverage for the same
date and time of the observations (10:00 UT). Note the excellent
agreement between model and observations. Red: Modeled spectral
albedo for 2003 December 7 at 21:00 UT. For 2003 December 7, a
strong vegetation red edge (starting around 700\,$nm$) is visible.
This is due to the large relative percentage of cloud-free vegetated
area in the Earth region contributing to the earthshine.}
\label{fig6}
\end{figure}

\begin{figure}[h]
\plottwo{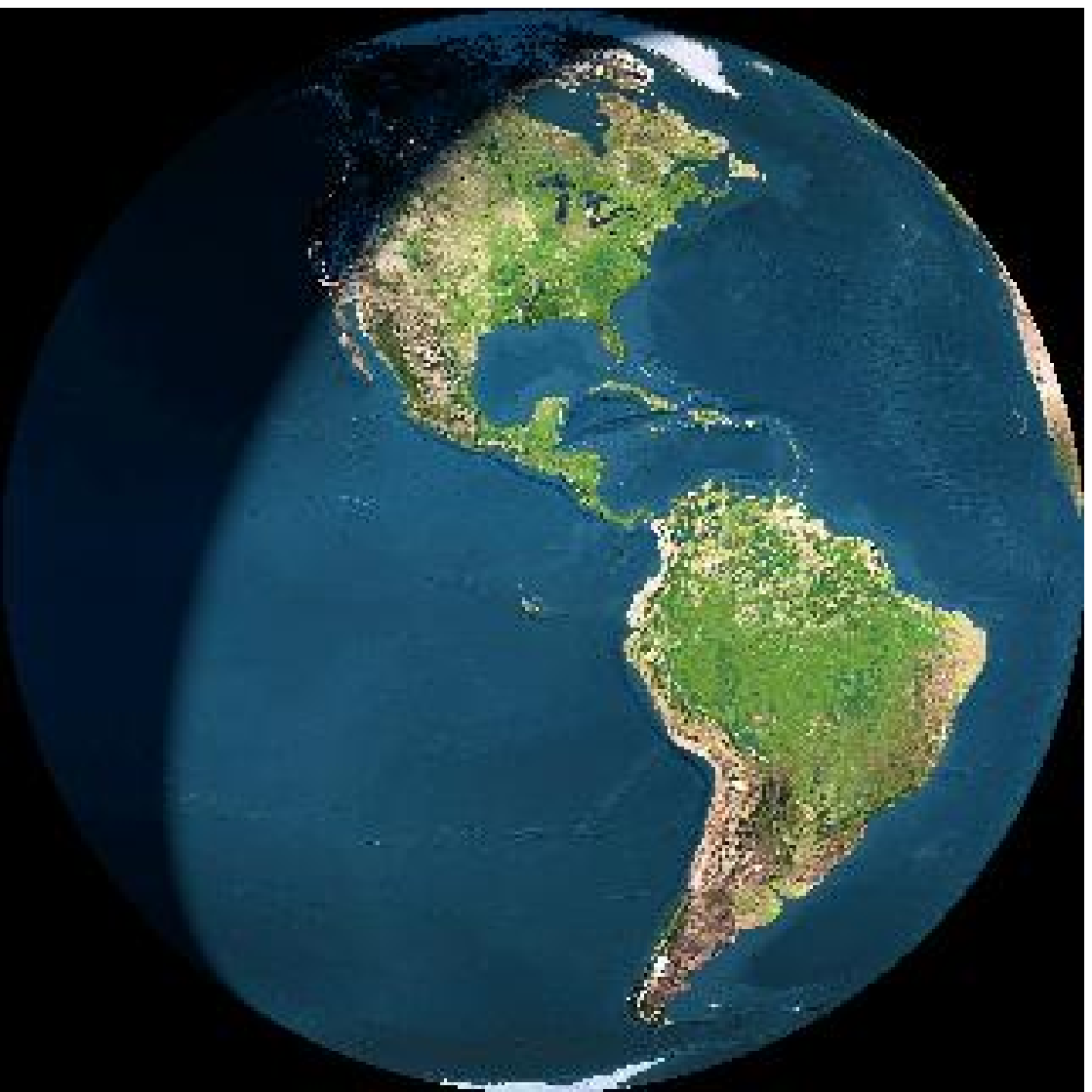}{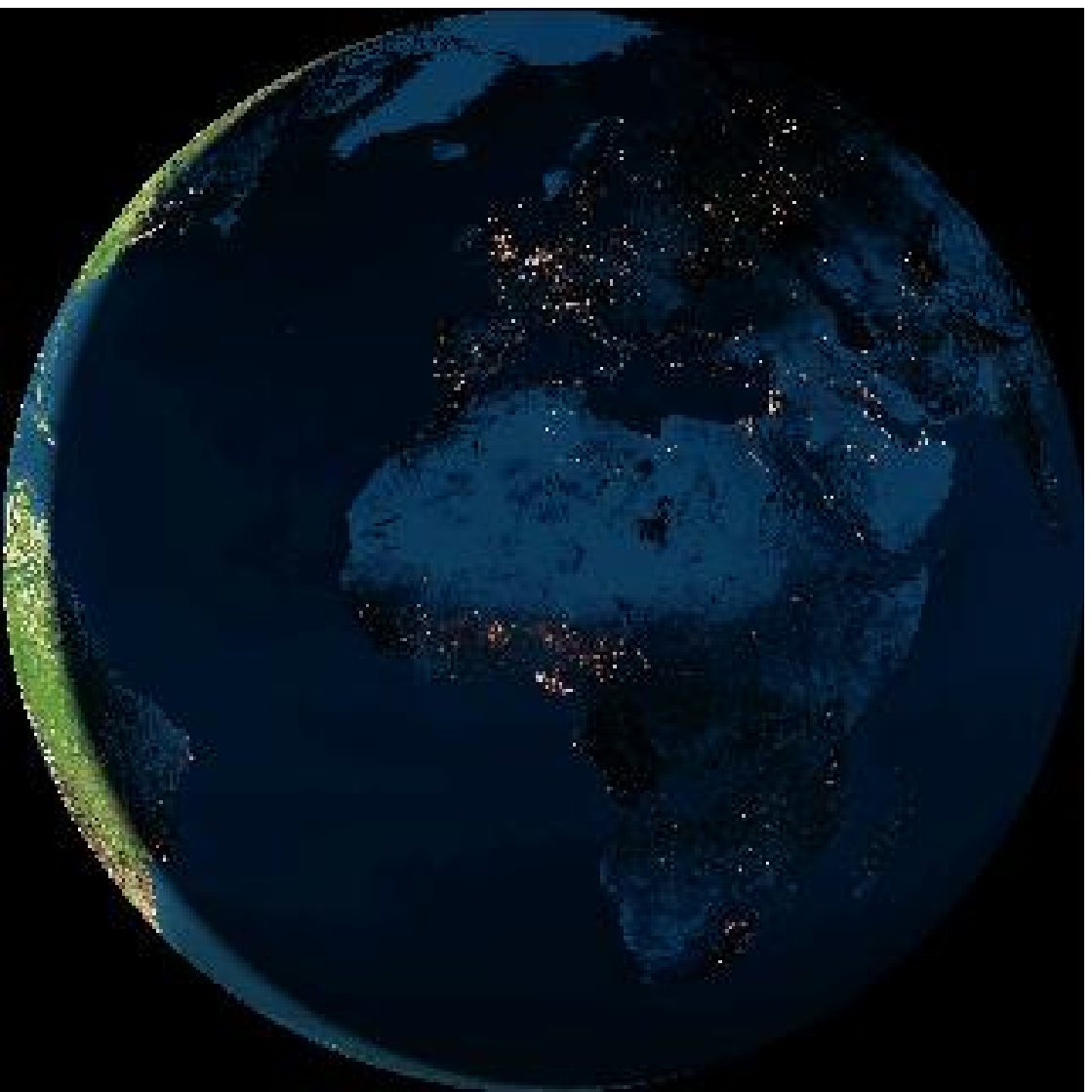} \caption{The
earthshine-contributing area as viewed from the Moon.  
Left: For 2003 November 19 at 10:00 UT (black data and green model in
Figure~\ref{fig6}), with lunar phase $+117^{o}$. 
Right: for 2003 December 7 at 21:00 UT (red
model in Figure~\ref{fig6}), with lunar phase $-25^{o}$. In the
left panel, large vegetated and un-vegetated areas are visible,
while in the right, the Earth is seen as a thin crescent and a large
proportion of the sunlit area is covered by vegetation. Note that
both images are very misleading, because the cloud cover is missing,
which will alter the scenery. The observations studied in this paper
for 2003 November 19 show only a very weak red edge despite the
Amazonian rainforest being in view. This is because in the left 
panel the total cloud-free vegetated area, projected in the Moon's
direction, is 15\% of the total sunlit area, while in the right
panel it is 48\%.} \label{twobarrel}
\end{figure}

The top panel in Figure~\ref{fig5} shows the 1-hour resolution
cloud-free vegetated area contributing to the earthshine for year
2003. Note that this contribution is in earthshine-contributing
surface percentage (weighted by the solar and lunar cosines at each
point of the Earth), but the real contribution to the earthshine
spectrum needs to be further weighted by the wavelength-dependent
albedo of each Earth scene component (clouds, deserts, etc...). For
most of the year, the vegetation's contribution is close to that
during 2003 November 19 for which ground-based earthshine
observations are available. However, the contribution peaks at
certain times, when the Earth as seen from the Moon is a thin
crescent, and the earthshine-contributing area is small. This is
illustrated in the lower panel of Figure~\ref{fig5}, where the daily
cycle of the vegetation visibility from the Moon is plotted for
three different days. This daily cycle depends on both the lunar
phase and the specific cloud distribution for that day. Note the
large vegetation contribution at certain times of the day 2002
December 19 reaching almost 50\%. At these times the red edge would
be easily detectable in the global spectrum, but the Earth's
crescent is so small that the flux in the observer's direction (in
our case, the Moon) is close to zero.

Thus, to detect vegetation one needs to find a compromise between
the Earth's scattering area, given by the projection of the total
sunlit planetary surface in the observer's direction, and the
maximum percentage of cloud-free vegetated area on that projection.
Analysis of 2003 cloud data indicates that during 70 hours (almost 3
full days), the scattering cross-section of the Earth was between
5-15\% of the maximum possible area (half of the planet), while
cloud-free vegetated areas were higher than 25\% of the contributing
area. Most of these dates occurred near, but not during, the full
Moon. Empirically, we have determined that a 25-30\%, or upward,
cloud-free vegetated area is enough to enable an unambiguous red
edge detection, but the particular detection threshold for a given
day will vary with cloud and surface type distributions in the
remaining area.

The earthshine spectrum for 2003 November 19 as observed from
Palomar Observatory (Monta\~n\'es-Rodr\'iguez et al. 2005a) is
plotted in Figure~\ref{fig6}. Also plotted is the modeled earthshine
spectrum using the ISCCP cloud distribution for the same date and
times. As shown in Section~3, model and observations agree well in
their overall shape and spectral features, but the vegetation red
edge is weak/missing in both because the signal is blocked by the
cloud cover. Also plotted is our modeled spectrum using the cloud
percentages for 2003 December 7 at 21:00 UT, a time when satellite
data indicate a cloud-free vegetated area close to 30\%
(Figure~\ref{fig5}). The model shows that the vegetation's red edge
is easily detectable at such times. Unfortunately observations for
dates like 2003 December 7 are impossible from ground-based
observatories, because at these times the Moon is almost full and
the earthshine is swamped by atmospheric light scattering. However,
our results indicate that if the reflected earthshine spectrum in
the dark side of the Moon were to be measured from satellite
platforms at lunar phases close to full Moon ($7^{o}-27^{o}$), a
clear vegetation signal would be detected.

\section{The Earthshine-Exoplanets connection}

\begin{figure*}
\epsscale{1.5} \plotone{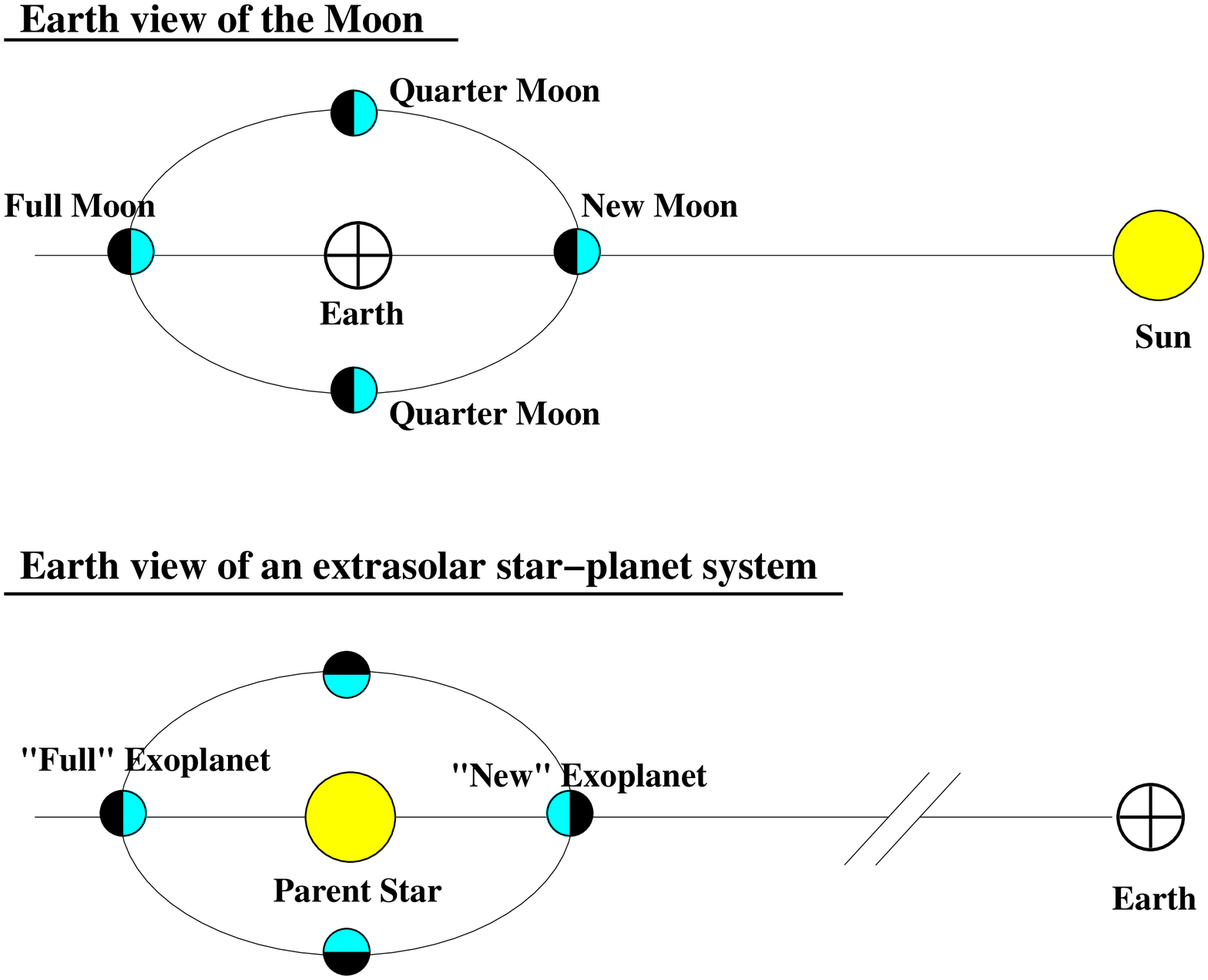} \caption{A
not-to-scale cartoon illustrating the similarity of observing the
earthshine at different lunar phases and observing an extrasolar
planet as it orbits its star.  Top: As the Moon rotates around the
Earth, it waxes and wanes from full Moon to new Moon. The phases of
the Earth and Moon are supplementary.  Thus, at new Moon, the area
contributing to the earthshine is half of the planet.  In other
words, an observer on the Moon would see a ``full'' Earth. Similarly
at full Moon, the earthshine contributing area is nearly zero.
Bottom:  As the exoplanet revolves around its star, a Earth-bound
observer would see the planet's scattering area waning and waxing.
For simplicity, in the cartoon the angle between the observer's line
of sight and the planet's ecliptic plane is set to be zero, but the
exoplanet's observable phase range from Earth will vary greatly
depending on this angle. Vegetation detection on the exoplanet will
become easier near ``new" (from the point of view of the Earth)
exoplanet phases, during glint scattering from heavily vegetated
area(s). However, there is a trade-off here because very close to
``new", it will be even more difficult to distinguish the exoplanet
from its star.} \label{fig7}
\end{figure*}

The earthshine case, i.e., the Moon's monthly evolving view of the
Earth, has a counterpart in the annual cycle of an extrasolar planet
around its parent star. Viewed from the distance, if our
observational line of sight is reasonably close to the planet's
ecliptic plane, the planet will offer waning and waxing phases with
its translation movement. This is illustrated in Figure~\ref{fig7}

There is a certain degree of uncertainty in applying our knowledge
based on Earth standards to extrasolar planets, but taking the
Earth's example, in order to sustain plant life, it is likely that
the planet needs substantial amounts of liquid water, i.e. oceans,
covering relatively large areas of its surface. It is also likely
that a significant amount of water vapor (clouds) is present, and
even tectonic activity (continents), to ensure climate stability,
may be necessary to harbor life (Kasting et al. 1984). Ignoring the
phytoplankton contribution, which has a lower albedo than leaves due
to its water cover (Gower et al. 2004), vegetation will be limited
to land areas. Thus, as the planet orbits around its star and
rotates, the vegetation signal should become more dominant at
certain periods of time as the planet approaches its orbital point
closer to the observer. Based on our above calculations, the
vegetation's red edge could be detectable in an earth-like planet
during 11\% of the planet's orbital period (40 days a year for
Earth), during an undetermined number of hours each day (typically
2-3 hours for Earth).

However, despite the advantageous vegetation signal, there are some
disadvantages to the observations of extrasolar planets at such
times of their orbits. When the planet is in the desired phase
range, typically only about 10\% of the planet's sunlit surface is
visible to the observer, which means that the planet's signal will
be faintest compared to its parent star. Considering the Earth as
roughly Lambertian (Qiu et al. 2003), the instrumental detection
power would need to increase between one and two orders of
magnitude, in order to retrieve the same spectral quality than for
the same planet viewed at phase $0^{o}$ (`full' planet).

Of course, for Earth-like extrasolar planets, the issue has the
additional complication of the strong signal from the star, which
would be more effectively nulled the greater the angular separation
of the planet and its star. If the extrasolar planet is angularly
close to the star, it is difficult to null-out the star (or
otherwise remove its spectral signature) without nulling out the
planet. This angular distance decreases as the planet reaches toward
``new", depending on the tilt angle between the observer and the
exoplanet's ecliptic plane. A small increase in this tilt angle will
also increase the angular separation between the exoplanet and the
parent star, but too much of an increase will reduce the observable
phase range of the exoplanet and eliminate those we are interested
in. Thus, there is a trade-off here to obtain observations at an
angular separation large enough to preserve the planet's signal, but
small enough to benefit from a rough geographical resolution that
allows the detection of vegetation. It is probable that none of the
missions currently on design phase will be able to retrieve this
measurements.


Through this manuscript, we have discussed plant detection in
extrasolar planets based on chlorophyll pigment properties typical
of Earth plants, but the universality of chlorophyll is still an
open question (Wolstencroft \& Raven 2002). The same detection
criteria, however, might apply for different pigmentation or
equivalent red edge signatures, characteristic of the plant type
that might exist in an extrasolar planet, provided that these
features do not coincide with spectral regions with strong
atmospheric absorption. Segura et al. (2005) and Tinetti et al. (2006b) 
have recently published studies about the possibility of an 
``equivalent" vegetation red edge on planets around M stars.

\section{Conclusions}

In our analysis of earthshine observations from 19 November 2003, we
find an increase in spectral reflectance, between 680 and 740\,$nm$
of the order of 0.005, which is highly correlated with the evolution
of the cloud-free vegetated areas contributing to the earthshine.
The good agreement between the evolution of the red-edge slope for
both models and observations with the total cloud-free vegetated
areas, might not be enough to confirm the detection of vegetation in
globally-integrated Earth's spectra. However, the 0.93 correlation
between the fixed cloud contribution models and the evolution of the
cloud-free vegetated area confirms it. Nevertheless, the change in
reflectance in this spectral region can only be related to
vegetation by analyzing the real distribution of land, oceans and
clouds during observations.

It is possible that an Earth-size extrasolar planet may have a
drastically different cloud cover system that yields a stronger
globally integrated vegetation signal. However, for an Earth-like
planet, the impossibility of knowing beforehand the cloud
distribution contributing to the planetary reflectance would seem to
make it very difficult to unambiguously detect the red edge
signature of vegetation.

We have also made use of the earthshine models and long-term
terrestrial cloud data to simulate the evolution of the red edge
signature with time, and we make the analogy between earthshine and
extrasolar planets; we demonstrate that there are days on Earth with
strong red edge signals, unlike the day discussed here, and that,
under certain orbital conditions, for a terrestrial-like planet
there are optimal times for detecting the red edge.

In summary, we have shown that the detection of vegetation on Earth
using globally-integrated spectral measurements is generally very
difficult, but becomes feasible under the appropriate geometrical
conditions. Similarly, for an Earth-like extrasolar planet, the
vegetation signature will become detectable during certain times of
its orbit. The excellent agreement of our earthshine observations
with modeled spectra of the Earth using real cloud distributions for
the same dates and times, gives us a strong confidence in our
results, which provide a reference for future space missions aiming
at the detection of such vegetation signatures in extrasolar
planets. In this case, observations will require extremely sensitive
instrumentation, still outside the scope of current design-phase
missions, but acquiring the necessary technology seems to be only a
matter of time.

\acknowledgments

This research was supported by the 2002 Ernest F. Fullam Award of 
the Dudley Observatory and by a grant from NASA (NASA-NNG04GN09G).

\end{document}